\documentclass[12pt]{iopart}
\usepackage{iopams}
\expandafter\let\csname equation*\endcsname\relax
\expandafter\let\csname endequation*\endcsname\relax
\usepackage{amsmath}
\usepackage{amssymb}
\usepackage{mathrsfs}

\newcommand{\mathsfbi}[1]{\boldsymbol{\mathsf{#1}}}
\begin{document}

\title{Field Theory of the Eulerian Perfect Fluid}

\author{Taketo Ariki${}^1$ \& Pablo A. Morales${}^2$}

\address{${}^1$Institute of Industrial Science, The University of Tokyo, 4-6-1 Komaba, Meguro-ku, Tokyo 153-8505, Japan\\
${}^2$Department of Physics, The University of Tokyo, 7-3-1 Hongo, Bunkyo-ku, Tokyo 113-0033, Japan}
\ead{${}^1$ariki@fluid.cse.nagoya-u.ac.jp, ${}^2$pablom@nt.phys.s.u-tokyo.ac.jp}

\vspace{10pt}
\begin{indented}
\item[]April 2017
\end{indented}

\begin{abstract}
The Eulerian perfect-fluid theory is reformulated from its action principle in a pure field-theoretic manner. Conservation of the convective current is no longer imposed by Lin's constraints, but rather adopted as the central idea of the theory. Our formulation, for the first time, successfully reduces redundant degrees of freedom promoting one half of the Clebsch variables as the true dynamical fields. Interactions on these fields allow for the exchange of the convective current of quantities such as mass and charge, which are uniformly understood as the breaking of the underlying symmetry of the force-free fluid. The Clebsch fields play the essential role in the exchange of angular momentum with the force field producing vorticity.
\end{abstract}

%
%
%
%
%

\section{Introduction}
The local nature of the laws governing fluid motion demand a field theoretical formulation. While the Lagrangian-coordinate framework focuses on fluid particle dynamics, its Eulerian approach describes the space-time behavior of fluid properties, representing a prototypical example of a classical field theory. Yet, in spite of its long history, fluid mechanics still lacks of a complete action principle acceptable as a field theory. Incompleteness arises in its description of the rotational flow; the action in the Eulerian description requires the \emph{Clebsch parametrization} \cite{C59, Bateman29, Eckart38, Danzig39, Eckart60, Lin, SW67, Sch70} of the fluid velocity $\mathbf{v}$, represented as a linear combination of scalar gradients 
\begin{equation}
\mathbf{v}=-\boldsymbol{\nabla} \phi -\frac{1}{\rho}s\boldsymbol{\nabla} \psi-\frac{1}{\rho}\beta\boldsymbol{\nabla}\alpha,
\label{Clebsch parametrization}
\end{equation}
where $\phi$ is the velocity potential, $s$ is the entropy density, $\psi$ is the \emph{thermasy} \cite{Danzig39} and $\rho$ is the mass density. The remaining $\alpha$ and $\beta$ are \emph{the Clebsch potentials} which are essential for rotational flow \cite{alphabeta}. This parametrization closely relates to the \emph{Lin constraints} at the Lagrangian density \cite{Lin}: 
\begin{equation}
\begin{split}
\mathcal{L}
=\mathcal{L}_0+\rho\mathfrak{D}\phi+s\mathfrak{D}\psi+\beta\mathfrak{D}\alpha,
\end{split}
\label{SW L}
\end{equation}
where the last three terms are the Lin constraints ($\mathfrak{D} (\equiv \partial_t+v^j\partial_j)$ is the convective derivative). The free part $\mathcal{L}_0\equiv \frac{1}{2}\rho v^jv_j-U(\rho,s)$ ($U$; internal energy) is constrained by conservation of $\rho$, $s$, and $\beta$ \cite{Eckart38, Lin, SW67, Sch70, Jackiw00, Kambe08b, FF10}. In other words, conventional action takes $\mathcal{L}_0$ as its core regarding the fluid as a particle assembly, while the convection of $\rho$, $s$ and $\beta$ are added as subsidiary constraints. In spite of its field-theoretic setup in coordinates, the conventional approach still relies on constructions built upon the particle substratum, and thus the Lin constraint unavoidable. Furthermore, the parametrization introduces several aspects around the potential variables themselves that are yet to be understood. For instance, ($I$) the Clebsh parametrization includes extra degrees of freedom of the Lagrange multipliers ($\phi,\ \psi$, and $\alpha$) of added constraints, which Schutz (1970) expresses as ``too many" and pointed out the necessity of action principle with a minimum number of variables \cite{Sch70}. ($I\!I$) Physical interpretation of the Clebsch potentials themselves are still controversial; some identify them as the Lagrangian-coordinate variables \cite{Lin, SW67, FF10}, some relate them with the Chern-Simons theory \cite{Jackiw00, Kambe08b}. Yet, no conclusive agreement on the nature of the potentials has been reached. In any event, any candidate formulation of the action principle in fluid mechanics should naturally account for the points described above. Perhaps, implying that the very basic set of ideas defining what we understand as fluid should be reconsidered.

Indeed, in revisiting the fundamental concepts of its field theoretical formulation, fluid dynamics has benefited from modern ideas developed in gauge theories. Within the Lagrangian coordinate approach, \cite{Nicolis} addresses the infrared dynamics of non-disipative fluid recasted in the language of effective field theory from symmetry considerations, allowing to explore viscous correction by means of its derivative expansion or anomalous hydrodynamics \cite{DTSon} among others. In a rather mixed framework, in \cite{Bis03,Jackiw04,Jackiw00,Jackiw04b} the Clebsch parametrization is realized by the introduction of a $SU(n)$ group element $g$ as potential variables. These novel approaches allowed the study of phenomena such as fluid with non-Abelian charges or spin-orbit interaction \cite{Nair14}, we refer the reader to \cite{Nair16} for a recent review. In particular, a through description of non-Abelian fluids can have far-reaching consequences in our understanding at the early stages of relativistic heavy ion collisions. An inclusive treatment dealing with interactions becomes crucial, and although appealing, their focus on single particle dynamics brings about some critical caveats; it still relies on a classical-particle substratum and an incomplete canonical formalism based on artificial symplectic structure. The fact that the dynamics governing fluid motion can be obtained from increasingly general principles is a reflection of the deeper and universal character of fluid theory.

In this paper, we go a step further and claim that a continuum's convection property is the core concept behind the field-theoretic formulation of fluid dynamics; we depart from the usual particle-based formulation and choose the convection effect as the guiding principle of fluid field theory, now for the first time, free of the Lin constraints. Reduction of such constraints naturally removes the redundant degrees of freedom from action, successfully minimizing the number of potentials. The physical picture provided by our formulation promotes the Clebsch variables as dynamical fields instead of a mere clever parametrization for the vector velocity. We discuss interactions and in the context of gauge symmetry apply to the case of the non-Abelian charged fluid. Our findings reveal not only assumptions disscused above can be avoided but a minimal number of fields are enough to capture dynamics of the non-Abelian fluid. We will show that, in such field-theoretic framework, the Clebsch potentials plays essential role in coupling with the force field, which allows the exchange of angular momentum with the external system.

\section{Formulation}
\subsection{Action of convection}\label{Action of convection}
Unlike the other physical fields, a fluid (or, generally speaking, continuum) is based on the velocity field by which various physical properties are transported; mass, electric charge, and other thermodynamic quantities are convected by this vector field with their current in the form of $\rho_{\mathrm{c}}\mathbf{v}$, where $\rho_{\mathrm{c}}$ is the density of some convective quantity; in general, convection is expressed in the form
\begin{equation}
\partial_t\rho_\mathrm{c}+\partial_j(\rho_\mathrm{c}v^j)=\mathrm{source\ term},
\label{convection eq.}
\end{equation}
whose conservation holds in the absence of the source. In this section we search for the most general Lagrangian $\mathcal{L}_\mathrm{c}$ yielding Eq. (\ref{convection eq.}) as the Euler-Lagrange equation from its internal symmetry. This means that $\rho_\mathrm{c}$ (a scalar) is obtained as a Noether charge, so $\mathcal{L}_\mathrm{c}$ should have a symmetry under a 1-dimensional transformation group; the associated symmetry is the shift of a scalar field, i.e. $\phi_\mathrm{c}\mapsto\phi_\mathrm{c}+const.$ (cyclic variable), and thus, we may at least write $\mathcal{L}_\mathrm{c}(\partial_t\phi_{\mathrm{c}},\partial_i\phi_\mathrm{c})$. Under a general-coordinate transformation in space $\{\mathbf{x}\}\mapsto\{\tilde{\mathbf{x}}\}$, $\partial_t\phi\mapsto\partial_t\phi+x^{\tilde{a}}{}_{,t}\partial_{\tilde{a}}\phi$, so $\partial_t\phi$ acquires 3D diffeomorphism invariance if accompanied by $+v^i\partial_i\phi$, where a vector $\mathbf{v}$ transforms as $v^{\tilde{a}}=x^{\tilde{a}}{}_{,i}v^i+x^{\tilde{a}}{}_{,t}$ which allows us to identify $\mathbf{v}$ as the velocity field \cite{A15}. Using the invariant $\mathfrak{D}\phi_\mathrm{c}\equiv(\partial_t +v^i\partial_i)\phi_\mathrm{c}$, we naturally obtain the Lagrangian density $\mathcal{L}_\mathrm{c}(\mathfrak{D}\phi_\mathrm{c})$. Indeed, in this way
\begin{subequations}
\begin{align}					
&\frac{\delta S_\mathrm{c}}{\delta \phi_\mathrm{c}}=-\partial_t \frac{\partial \mathcal{L}_\mathrm{c}}{\partial(\mathfrak{D}\phi_\mathrm{c})}
-\partial_j\left(\frac{\partial \mathcal{L}_\mathrm{c}}{\partial(\mathfrak{D}\phi_\mathrm{c})} v^j\right)=0,
\label{phi eq.}\\
&\frac{\delta S_\mathrm{c}}{\delta v^j}
=\frac{\partial \mathcal{L}_\mathrm{c}}{\partial(\mathfrak{D}\phi_\mathrm{c})}\phi_{\mathrm{c},j}=0,
\label{v eq.}
\end{align}
\end{subequations}
where $\partial \mathcal{L}_\mathrm{c}/\partial (\mathfrak{D}\phi_\mathrm{c})$ serves as the convective quantity $\rho_\mathrm{c}$ in comparison with Eq. (\ref{convection eq.}). The relation with the Lin's constraint in Eq. (\ref{SW L}) will be clarified in the Hamiltonian formalism. The canonical conjugate of $\phi_\mathrm{c}$ is $\pi_\mathrm{c}=\partial \mathcal{L}_\mathrm{c}/\partial (\partial_t\phi_\mathrm{c})=\partial \mathcal{L}_\mathrm{c}/\partial (\mathfrak{D}\phi_\mathrm{c})(\equiv f_\mathrm{c}(\mathfrak{D}\phi_\mathrm{c}))=\rho_\mathrm{c}$, where we shall choose the functional form of $\mathcal{L}_\mathrm{c}$ so that the inverse function $F_\mathrm{c}(\equiv f_\mathrm{c}^{-1}); \rho_\mathrm{c}\mapsto \mathfrak{D} \phi_\mathrm{c}$ exists, namely, $\mathfrak{D}\phi_\mathrm{c}=F_\mathrm{c}(\rho_\mathrm{c})$. Then the Hamiltonian density reads:
\begin{equation}
\mathcal{H}_\mathrm{c}=\rho_\mathrm{c}\partial_t \phi_\mathrm{c}-\mathcal{L}_\mathrm{c}\circ F_\mathrm{c}(\rho_\mathrm{c})
=-\rho_\mathrm{c}v^j\phi_{,j}+\int^{\rho_\mathrm{c}} F_\mathrm{c}(\xi)\mathrm{d}\xi,
\end{equation}
then 
\begin{equation}
\begin{split}
S_\mathrm{c}=\int\mathrm{d}t\int\mathrm{d}^3x
\left(\rho_\mathrm{c}\mathfrak{D}\phi_\mathrm{c}-\int^{\rho_\mathrm{c}} F_\mathrm{c}(\xi)\mathrm{d}\xi\right),
\end{split}
\label{H of convection}
\end{equation}
which adds a non-trivial modification to the simple Lin constraint $S_\mathrm{Lin}=\int\mathrm{d}t\int\mathrm{d}^3x\rho_\mathrm{c}\mathfrak{D}\phi_\mathrm{c}$ (see \ref{derivation of Hc} for its derivation). Here we soon recognize that the potential pair $(\phi_\mathrm{c},\rho_\mathrm{c})$ in the Lin constraint is reproduced as the canonical-conjugate pair. One should not, however, mistake $S_\mathrm{c}[\phi_\mathrm{c},\mathbf{v}]$ as the phase-space counterpart of $S_\mathrm{Lin}[\phi_\mathrm{c},\rho_\mathrm{c}\mathbf{v}]$; \emph{the Lin constraint alone cannot be a complete canonical action of $(\phi_\mathrm{c},\rho_\mathrm{c})$}, precisely due to the presence of $\int^{\rho_\mathrm{c}} F_\mathrm{c}(\xi)\mathrm{d}\xi$ in Eq. (\ref{H of convection}). Although both $S_\mathrm{c}$ and $S_\mathrm{Lin}$ result in the same convective equation ($\partial_t\rho_\mathrm{c}+(\rho_\mathrm{c}v^j)_{,j}=0$), the equation for $\phi_\mathrm{c}$ changes due to this modification; namely $S_\mathrm{c}$ yields $\mathfrak{D}\phi_\mathrm{c}=F_\mathrm{c}(\rho_\mathrm{c})$ while $S_\mathrm{Lin}$ yields $\mathfrak{D}\phi_\mathrm{c}=0$. Thus, the proposed $S_\mathrm{c}[\phi_\mathrm{c},\mathbf{v}]$ successfully reduces one degree of freedom from $S_\mathrm{Lin}[\rho_\mathrm{c},\phi_\mathrm{c},\mathbf{v}]$ by modifying a non-observed $\phi_\mathrm{c}$ equation. Therefore, the phase space $\{\phi_\mathrm{c},\rho_\mathrm{c}\}$ of conventional Lin constraint based formulations cannot be reduced to $\{\phi_\mathrm{c}\}$. 

Following the same strategy, the fluid model of Eq. (\ref{SW L}) written by 6 scalars will be reformulated by only 3 potentials (see \S\ref{Hamiltonian formalism}). If one does not consider the entropy, the minimal model of rotational-barotropic fluid is written by only 2 potentials (see \S\ref{Kernel formalism}). Although some pioneers in \cite{SW67, Sch70, Jackiw00} have found that the potentials in the Clebsch parametrization form canonical pairs, they really could not reduce the phase space $\{\phi,\psi,\alpha;\rho,s,\beta\}$ down to the configuration space $\{\phi,\psi,\alpha\}$. In order to obtain the real action in the configuration space, we shall abandon the Lin constraint of the standard formalisms \cite{Eckart38, Lin, SW67, Sch70, Kambe08b, FF10, Jackiw00, Jackiw04, Jackiw04b}.

\subsection{Kernel formalism}\label{Kernel formalism}
In this section, following the logic behind the previously obtained action of convection $S_\mathrm{c}$, we establish the element and formalism that will lead us to the most general fluid Lagrangian density. We start with $\phi$ and $\mathbf{v}$ as dynamical variables. As seen in Eq. (\ref{v eq.}), Lagrangian density $\mathcal{L}_\mathcal{F}(\mathfrak{D}\phi)$ alone does not contain the dynamics of $\mathbf{v}$. Thus, dependence on the kinetic term $\mathcal{V}=\frac{1}{2}v^av_a$ is imposed, $\mathcal{L}_\mathcal{F}(\mathfrak{D}\phi,\mathcal{V})$. Eq. (\ref{v eq.}) changes accordingly
\begin{equation}
\frac{\delta S}{\delta v^j}
=\frac{\partial \mathcal{L}_\mathcal{F}}{\partial(\mathfrak{D}\phi)}\phi_{,j}+\frac{\partial \mathcal{L}_\mathcal{F}}{\partial\mathcal{V}}v_j=0,
\label{v eq. +}
\end{equation}
where $\phi$ serves as a potential function. This becomes exactly the potential relation $\mathbf{v}=-\boldsymbol{\nabla} \phi$ when the Langrangian density shares the same dependance in both $\mathcal{V}$ and $\mathfrak{D}\phi$, i.e. $\partial\mathcal{L}_\mathcal{F}/\partial\mathcal{V}=\partial \mathcal{L}_\mathcal{F}/\partial(\mathfrak{D}\phi)$. One could relax this condition so that they are proportional to each other, in this case by rescaling $\phi$ in the equations above we keep the potential relation intact. Alternatively, we introduce a new quantity $\tilde{\mathfrak{D}}\phi$ as
\begin{equation}
\tilde{\mathfrak{D}}\phi=\mathfrak{D}\phi+\frac{1}{2}Mv^jv_j,
\label{kernel 1}
\end{equation}
and rewrite the Lagrangian density as $\mathcal{L}_\mathcal{F}(\tilde{\mathfrak{D}}\phi)$, with $M$ a constant. $\tilde{\mathfrak{D}}\phi$, to be referred as the \emph{kernel} hereafter, plays the central role in the later discussions. The corresponding Euler-Lagrange equations read
\begin{subequations}
\begin{align}
&\frac{\delta S}{\delta \phi}=-\partial_t \frac{\partial \mathcal{L}_\mathcal{F}}{\partial(\tilde{\mathfrak{D}}\phi)}
-\left(\frac{\partial \mathcal{L}_\mathcal{F}}{\partial(\tilde{\mathfrak{D}}\phi)} v^j\right){}_{,j}=0,\label{phi eq. 2}\\
&\frac{\delta S}{\delta v^j}
=\frac{\partial \mathcal{L}_\mathcal{F}}{\partial(\tilde{\mathfrak{D}}\phi)}(\phi_{,j}+M v_j)=0,
\label{v eq. 2}
\end{align}
\end{subequations}
where convection of $\partial\mathcal{L}_\mathcal{F}/\partial \tilde{\mathfrak{D}}\phi$ and the potential relation are obtained. Note that Lagrangian density $\mathcal{L}_\mathcal{F}(\tilde{\mathfrak{D}}\phi)$ with the kernel (\ref{kernel 1}) provides a minimal model for the perfect fluid. The spatial gradient of $\mathcal{L}_\mathcal{F}(\tilde{\mathfrak{D}}\phi)$ yields
\begin{equation*}
\begin{split}
\partial_i\mathcal{L}_\mathcal{F}(\tilde{\mathfrak{D}}\phi)
&=\frac{\partial \mathcal{L}_\mathcal{F}}{\partial(\tilde{\mathfrak{D}}\phi)}(\phi_{,ti}+v^j{}_{,i}\phi_{,j}+v^j\phi_{,ji}+Mv^j{}_{,i}v_j)\\
&=-M\frac{\partial \mathcal{L}_\mathcal{F}}{\partial(\tilde{\mathfrak{D}}\phi)}
(v_{i,t}+v_{i,j}v^j)
\end{split}
\end{equation*}
\begin{equation}
\Leftrightarrow
M\frac{\partial \mathcal{L}_\mathcal{F}}{\partial(\tilde{\mathfrak{D}}\phi)}
(v_{i,t}+v_{i,j}v^j)=-{\mathcal{L}_\mathcal{F}}_{,i},
\label{Euler eq.}
\end{equation}
where we used Eq. (\ref{v eq. 2}). Eq. (\ref{Euler eq.}) is the Euler equation where $M\partial\mathcal{L}_\mathcal{F}/\partial\tilde{\mathfrak{D}}\phi$ and $\mathcal{L}_\mathcal{F}$ behave as the mass density and the pressure respectively. The pressure-mass relation is better appreciated once rewritten in terms of the canonical momentum of $\phi$, $\sigma$ defined as
\begin{equation*}
\sigma = \frac{\partial\mathcal{L}_\mathcal{F}}{\partial \phi_{,t}}
=\frac{\partial\mathcal{L}_\mathcal{F}}{\partial(\tilde{\mathfrak{D}} \phi)}
\left(\equiv f(\tilde{\mathfrak{D}} \phi)\right).
\end{equation*} 
Then, the constitutive relation between pressure $\mathcal{L}_\mathcal{F}$ and mass $M\sigma$ is given by $(\mathcal{L}_\mathcal{F}\circ F)(\sigma)$, where $F(\equiv f^{-1}); \sigma\mapsto\tilde{\mathfrak{D}}\phi$. Thus $\mathcal{L}_\mathcal{F}(\tilde{\mathfrak{D}}\phi)$ with the kernel (\ref{kernel 1}) describes an irrotational barotropic fluid whose pressure is determined only by mass density $M\sigma$.\\ 

In describing rotational flow, a reasonable guess would be to impose the Clebsch potential $\alpha$ into the Lagrangian as $\mathcal{L}_\mathcal{F}(\tilde{\mathfrak{D}}\phi,\mathfrak{D}\alpha)$, then $\mathbf{v}=-\boldsymbol{\nabla}\phi/M-\beta\boldsymbol{\nabla}\alpha/M\sigma$ ($\beta\equiv \partial \mathcal{L}_\mathcal{F}/\partial\,\mathfrak{D}\alpha$). However, such approach fails as the resulting pressure $\mathcal{L}_\mathcal{F}$ would depend on both $\sigma$ and $\beta$ which contradicts the nature of the Clebsch potentials. In other words $\tilde{\mathfrak{D}}\phi$ and $\alpha$ are not independant variables of the Lagrangian density meaning it can be rewritten as a composite function $\mathcal{L}_\mathcal{F}\circ A(\tilde{\mathfrak{D}}\phi,\mathfrak{D}\alpha)$, being $A(\tilde{\mathfrak{D}}\phi,\mathfrak{D}\alpha)$ some real function. The canonical momentum $\sigma$ now reads,
\begin{equation}
\sigma=\frac{\partial\mathcal{L}_\mathcal{F}}{\partial A}(A)\frac{\partial A}{\partial \tilde{\mathfrak{D}}\phi}(\tilde{\mathfrak{D}}\phi,\mathfrak{D}\alpha).
\end{equation}
Only when $\partial A/\partial \tilde{\mathfrak{D}}\phi$ becomes a constant, a function $F;\sigma\mapsto A$ exists so that the pressure $p=\mathcal{L}_\mathcal{F}\circ A\circ F(\sigma)$ is free from $\beta$. This requires $A\propto\tilde{\mathfrak{D}}\phi+C(\mathfrak{D}\alpha)$ with a real function $C(\mathfrak{D}\alpha)$. The demanded Lagrangian density is most simply written as $\mathcal{L}_\mathcal{F}(\tilde{\mathfrak{D}}\phi)$ with the modified kernel 
\begin{equation}
\tilde{\mathfrak{D}}\phi=\mathfrak{D}\phi+\frac{1}{2}Mv^jv_j+C(\mathfrak{D}\alpha),
\label{kernel 2}
\end{equation}
which gives the minimal model of rotational barotropic fluid, this time described only by 2 potentials.


The kernel of Eq. (\ref{kernel 2}) may be the most fundamental ingredient in the description of a fluid, since any extension $\mathcal{L}_\mathcal{F}(\tilde{\mathfrak{D}}\,{}^{{}_1}\!\phi,\tilde{\mathfrak{D}}\,{}^{{}_2}\!\phi,\cdots)$ containing arbitrary numbers of kernels will always yield the same fluid equations. This function constitutes the most general Lagrangian density for the perfect fluid, which may be rewritten in a more compact form $\mathcal{L}_\mathcal{F}(\tilde{\mathfrak{D}}\mathsfbi{\phi})$ by taking the kernel as a multiple-component object:
\begin{equation}
\tilde{\mathfrak{D}}\boldsymbol{\phi}=\mathfrak{D}\boldsymbol{\phi}+\frac{1}{2}\mathsfbi{M}v^jv_j+\mathsfbi{C}(\mathfrak{D}\mathsfbi{\alpha}).
\label{kernel 3}
\end{equation}
Here $\mathsfbi{\phi}$ and $\mathsfbi{\alpha}$ are $m$- and $n$-component real scalars respectively, $\mathsfbi{C}$ is a $m$-component real function of $\mathfrak{D}\mathsfbi{\alpha}$, $\mathsfbi{M}$ is $m$-component constant ($m\neq n$ in general):
\begin{equation*}
\begin{split}
&\mathsfbi{\phi}=({}^{{}_1}\!\phi,\,{}^{{}_2}\!\phi,\cdots,\,{}^{{}_m}\!\phi)^{{}_T},\ \ \mathsfbi{\alpha}=({}^{{}_1}\!\alpha,\ {}^{{}_2}\!\alpha,\cdots,\ {}^{{}_n}\!\alpha,)^{{}_T},\\
&\mathsfbi{C}=({}^{{}_1}\!C,\,{}^{{}_2}\!C,\cdots,\,{}^{{}_m}\!C)^{{}_T},\ \ 
\mathsfbi{M}=({}^{{}_1}\!M,\,{}^{{}_2}\!M,\cdots,\,{}^{{}_m}\!M)^{{}_T},
\end{split}
\end{equation*}
where $T$ attached on the right-top side means the transposition of matrix components. Hereafter, vectors such as $\mathsfbi{\phi}$ ($\mathsfbi{\phi}^{{}_T}$) are regarded as column (row) vectors, while the derivative operations such as $\partial/\partial \mathsfbi{\phi}$ ($\partial/\partial \mathsfbi{\phi}^{{}_T}$) as rows (columns). We have two canonical conjugates $\mathsfbi{\sigma}$ and $\mathsfbi{\beta}$ given by
\begin{subequations}
\begin{align*}
&\boldsymbol{\sigma}^{{}_T}=\frac{\partial\mathcal{L}_\mathcal{F}}{\partial \boldsymbol{\phi}_{,t}}
=\frac{\partial\mathcal{L}_\mathcal{F}}{\partial(\tilde{\mathfrak{D}}\boldsymbol{\phi})}
=\mathsfbi{f}^{{}_T}(\tilde{\mathfrak{D}}\mathsfbi{\phi}),\\
&\mathsfbi{\beta}^{{}_T}=\frac{\partial\mathcal{L}_\mathcal{F}}{\partial\mathsfbi{\alpha}_{,t}}
=\frac{\partial\mathcal{L}_\mathcal{F}}{\partial(\tilde{\mathfrak{D}}\boldsymbol{\phi})}
\frac{\partial\mathsfbi{C}}{\partial(\mathfrak{D}\boldsymbol{\alpha})}
=\mathsfbi{\sigma}^{{}_T}\mathsfbi{h}(\mathfrak{D}\mathsfbi{\alpha}), 
\end{align*}
\end{subequations}
where $\mathsfbi{h}(\equiv\partial\mathsfbi{C}/\partial \mathfrak{D}\mathsfbi{\alpha})$ is $m\times n$ matrix. The Euler-Lagrange equations are given by
\begin{subequations}
\begin{align}
&\frac{\delta S}{\delta\mathsfbi{\phi}^{{}_T}}=-\mathsfbi{\sigma}_{,t}-\left(\mathsfbi{\sigma} v^{i}\right){}_{,i}=0,
\label{phi eq. 3}\\
&\frac{\delta S}{\delta\mathsfbi{\alpha}^{{}_T}}=-\mathsfbi{\beta}_{,t}-\left(\mathsfbi{\beta} v^{i}\right){}_{,i}=0,\label{alpha eq. 3}\\
&\frac{\delta S}{\delta v^i}
=\mathsfbi{\sigma}^{{}_T}\mathsfbi{\phi}_{,i}
+\mathsfbi{\beta}^{{}_T}\mathsfbi{\alpha}_{,i}
+\mathsfbi{M}^{{}_T}\mathsfbi{\sigma} v_i=0,
\label{v eq. 3}
\end{align}
\end{subequations}
where Eq. (\ref{v eq. 3}) is the generalization of Eq. (\ref{Clebsch parametrization}). Taking the spatial derivative of $\mathcal{L}_\mathcal{F}(\tilde{\mathfrak{D}}\mathsfbi{\phi})$ and combined with Eq. (\ref{v eq. 3}) yield
\begin{equation}
\mathsfbi{M}^{{}_T}\mathsfbi{\sigma}\left(v_{i,t}+v_{i,j}v^j\right)=-\mathcal{L}_\mathcal{F}{}_{,i}.
\label{Euler eq. 2}
\end{equation}
Now $\mathsfbi{M}^{{}_T}\mathsfbi{\sigma}$ behaves as the mass density. Indeed, (\ref{phi eq. 3}) multiplied by $\mathsfbi{M}$ yields the conservation law as follows:
\begin{equation}
\partial_t (\mathsfbi{M}^{{}_T}\mathsfbi{\sigma})
+(\mathsfbi{M}^{{}_T}\mathsfbi{\sigma}v^j)_{,j}=0.
\label{mass cons}
\end{equation}
As evident from the Euler-Lagrange equations, the Lagrangian density $\mathcal{L}_\mathcal{F}(\tilde{\mathfrak{D}}\boldsymbol{\phi})$ has its foundation on the convection property, and not on momentum conservation. As a consequence, the Euler equation (\ref{Euler eq. 2}) and mass conservation (\ref{mass cons}) are automatically obeyed, in this sense they are \emph{universal} features of the fluid rather than conditions to be imposed. In other words, in deriving Eqs. (\ref{Euler eq. 2}) and (\ref{mass cons}) alone, the known thermodynamic variables and their relations are not necessary. Then, the perfect fluid is not only an idealized model of known materials, but rather a universal feature lying behind the convection. Needless to say, both hold for arbitrary values of $m$ and $n$, which can be thought as a discrete symmetry of our Lagrangian.

\subsection{Hamiltonian formalism}\label{Hamiltonian formalism}
Let us turn to the Hamiltonian formalism, here we can already see some clear differences between the conventional and present theories. The canonical Hamiltonian density reads,
\begin{equation}
\begin{split}
\mathcal{H}_\mathcal{F}
&=\mathsfbi{\sigma}^{{}_T}\mathsfbi{\phi}_{,t}
+\mathsfbi{\beta}^{{}_T}\mathsfbi{\alpha}_{,t}
-\mathcal{L}_\mathcal{F}\\
&=U(\mathsfbi{\sigma})+U_C(\mathsfbi{\beta},\mathsfbi{\sigma})
-\frac{1}{2}\mathsfbi{M}^{{}_T}\mathsfbi{\sigma}v_a v^a-v^j(\mathsfbi{\sigma}^{{}_T}\mathsfbi{\phi}_{,j}+\mathsfbi{\beta}^{{}_T}\mathsfbi{\alpha}_{,j}),
\end{split}
\label{H}
\end{equation}
where $U(\boldsymbol{\sigma})=\int^{\mathsfbi{\sigma}}\mathsfbi{F}^{{}_T}(\mathsfbi{\sigma}')\mathrm{d}\mathsfbi{\sigma}'$ ($\mathsfbi{F};\mathsfbi{\sigma}\mapsto \tilde{\mathfrak{D}}\mathsfbi{\phi}$ is the inverse of $\mathsfbi{f}$) is the internal energy, and $U_C(\boldsymbol{\beta},\boldsymbol{\sigma})=\int^{\mathsfbi{\beta}}\mathsfbi{H}^{{}_T}(\mathsfbi{\beta}',\mathsfbi{\sigma})\,\mathrm{d}\mathsfbi{\beta}'$ ($\mathsfbi{H}; \mathsfbi{\beta} \mapsto \mathfrak{D}\mathsfbi{\alpha}$ is obtained as the inverse of $\mathsfbi{h}$ for fixed $\mathsfbi{\sigma}$), both of which correspond to $\int^{\rho_\mathrm{c}}F_\mathrm{c}(\xi)\mathrm{d}\xi$ in Eq. (\ref{H of convection}). For a special case $(m,n)=(2,1)$ with $\mathsfbi{M}=(1,0)^{{}_T}$, Eq. (\ref{H}) is reduced to
\begin{equation}
\begin{split}
\mathcal{H}_\mathcal{F}
&=U(\rho,s)+U_C(\rho,s,\beta)-\frac{1}{2}\rho v_a v^a
-v^j(\rho\phi_{,j}+s\psi_{,j}+\beta\alpha_{,j}),
\end{split}
\label{H2}
\end{equation}
where $\mathsfbi{\phi}=(\phi,\psi)^{{}_T}$ and $\mathsfbi{\sigma}=(\rho,s)^{{}_T}$. Note that our derivation of the Hamiltonian (\ref{H2}) does not trivially follow from fluid energy $\mathcal{H}=\frac{1}{2}\rho v^2 +U$ nor it relies on given Poisson brackets as \cite{Jackiw04,Nair12,Nair16} do. In particular, $U_C(\rho,s,\beta)$ does not appear in the conventional formulation. And, although it does not affect the conservation of $\rho$, $s$, and $\beta$ \cite{UC}, modifies the dynamics of $\phi$, $\psi$, and $\alpha$:
\begin{subequations}
\begin{align}
\partial_t\phi&=\frac{\delta \mathscr{H}_\mathcal{F}}{\delta \rho}=\frac{\partial U(\rho,s)}{\partial \rho}+\frac{\partial U_C(\rho,s,\beta)}{\partial \rho}-\frac{1}{2}v^2-v^j\phi_{,j},\label{canonical phi}\\
\partial_t\psi&=\frac{\delta \mathscr{H}_\mathcal{F}}{\delta s}=T+\frac{\partial U_C(\rho,s,\beta)}{\partial s}-v^j\psi_{,j},\label{anomalous thermasy}\\
\partial_t\alpha&=\frac{\delta \mathscr{H}_\mathcal{F}}{\delta \beta}=\frac{\partial U_C(\rho,s,\beta)}{\partial \beta}-v^j\alpha_{,j},\label{anomalous alpha}
\end{align}
\end{subequations}
where $\mathscr{H}_\mathcal{F}(\equiv\int\mathcal{H}_\mathcal{F}\mathrm{d}^3x)$ is the total Hamiltonian, $T(\equiv\partial U/\partial s)$ is the temperature. Eqs. (\ref{anomalous thermasy}) and (\ref{anomalous alpha}) yield $\mathfrak{D}\psi-T=\partial U_C/\partial s$ and $\mathfrak{D}\alpha=\partial U_C/\partial \beta$, while the conventional approaches give $\mathfrak{D}\psi=T$ and $\mathfrak{D}\alpha=0$ (see Eq. (2.22) in \cite{Sch70}, Eqs. (32)-(33) of \cite{SW67}, and Eq. (7) of \cite{Bis03}, where they denote it as $\beta$ for our $\alpha$). Then our $\psi$ is no more the conventional thermasy but rather its extension. Also note that the presence of $U_C$ leads to the extension of the known Bernoulli theorem to the non-barotropic, rotational, and unsteady flow; $\rho\times$(\ref{canonical phi}) $+$ $s\times$ (\ref{anomalous thermasy}) $+$ $\beta\times$ (\ref{anomalous alpha}) yield
\begin{equation}
\begin{split}
\rho \phi_{,t}+s\psi_{,t}+\beta \alpha_{,t}-\frac{1}{2}\rho v^2
=\left(\rho\frac{\partial}{\partial\rho} +s\frac{\partial}{\partial s}+\beta\frac{\partial}{\partial\beta}\right)(U+U_C)(\rho,s,\beta),
\end{split}
\label{generalized Bernoulli eq.}
\end{equation}
which gives the first integral of the Euler equation by using $U_C$. Likewise, $U_C$ does not work as the usual potential energy, but as one sector of the total Hamiltonian generating the motions of $\phi$, $\psi$, and $\alpha$, which alters the physical interpretation of velocity potentials.

Presence of $U_C$ modifies the physical interpretation of $\mathcal{H}_\mathcal{F}$ in a non-trivial manner. Using the canonical equations, we soon realize $\mathcal{H}_\mathcal{F}=\frac{1}{2}\rho v^2+U+U_C$, so, in the phase-space trajectory, the Hamiltonian takes the value of the fluid energy plus the Clebsch-potential energy $U_C$. Thus the fluid-energy conservation does not trivially hold from the conservation of the total Hamiltonian. Considering $\partial_t U_C=\{U_C, \mathscr{H}_\mathcal{F}\}_\mathrm{PB}=-\partial_j(U_C v^j)$, meaning $U_C$ also forms convective current conserving in the whole space, we reach the conservation of $\frac{1}{2}\rho v^2+U$.

Finally, let us see the relation with Eq. (\ref{SW L}). In the canonical formalism, the Lagrangian density reads
\begin{equation}
\begin{split}
\mathcal{L}_\mathcal{F}&=\rho\phi_{,t}+s\psi_{,t}+\beta\alpha_{,t}-\mathcal{H}_\mathcal{F}\\
&=\frac{1}{2}\rho v^jv_j-U(\rho,s)-U_C(\rho,s,\beta)
+\rho\mathfrak{D}\phi+s\mathfrak{D}\psi+\beta\mathfrak{D}\alpha.
\end{split}
\label{SW L 2}
\end{equation}
Although (\ref{SW L 2}) looks similar to Eq. (\ref{SW L}), the presence of Clebsch-potential energy $U_C$ results in the different Euler-Lagrange equations. Also it is highly non-trivial that the last three terms at Eq. (\ref{SW L 2}) are not added constraints as in Eq. (\ref{SW L}) but naturally emerge through the Legendre transformation $\mathcal{L}_\mathcal{F}[\mathsfbi{\phi},\mathsfbi{\alpha}]\rightarrow\mathcal{H}_\mathcal{F}[\mathsfbi{\phi},\mathsfbi{\alpha};\mathsfbi{\sigma},\mathsfbi{\beta}]$ at Eq. (\ref{H}); namely these three are no more constraints. These difference make a critical gap between the present and the conventional in terms of the dimensional reduction. On the basis of the Lagrangian density Eq. (\ref{SW L}), due to $U_C$'s absence and different equations for potentials, the phase space $\{\phi,\psi,\alpha\ ;\ \rho,s,\beta\}$ cannot be reduced to the configuration space $\{\phi,\psi,\alpha\}$ like we just have. As pointed out in \S\ref{Action of convection}, the Lin's constraint and the conventional potential equations are to be discarded in order to obtain the true Lagrangian density formulated in the configuration space.


\section{Relativistic fluid}
\subsection{Relativistic kernel}
The relativistic-fluid theory follows from its relativistic kernel. For simplicity of discussions, we seek for the fluid action in flat-space $S=\int\mathcal{L}_\mathcal{F}(\tilde{\mathfrak{D}}\mathsfbi{\phi})\mathrm{d}^4x$ (we adopt metric signature: $g_{\mu\nu}=\textrm{diag}(1,-1,-1,-1)$). Replacing the non-relativistic velocity $\mathbf{v}$ with a four-dimensional time-like vector $\mathbf{w}$, the relativistic kernel is introduced as follows:
\begin{equation}
\tilde{\mathfrak{D}}\mathsfbi{\phi}=\mathfrak{D}\mathsfbi{\phi}
-\frac{1}{2}\mathsfbi{M}w_{\alpha}w^{\alpha}+\mathsfbi{C}(\mathfrak{D}\mathsfbi{\alpha}),
\label{kernel 4}
\end{equation}
where $\mathfrak{D}\equiv w^{\alpha}\partial_{\alpha}$. In the later discussions we employ the Greek alphabets for the indices of space-time coordinates. The Euler-Lagrange equations are given by
\begin{subequations}
\begin{align}
&\frac{\delta S}{\delta\mathsfbi{\phi}^{{}_T}}=-\left(\mathsfbi{\sigma} w^{\mu}\right){}_{,\mu}=0,\label{phi eq. 4}\\
&\frac{\delta S}{\delta \mathsfbi{\alpha}^{{}_T}}=-\left(\mathsfbi{\beta} w^{\mu}\right){}_{,\mu}=0,\label{alpha eq.4}\\
&\frac{\delta S}{\delta w^{\mu}}
=\mathsfbi{\sigma}^{{}_T}\mathsfbi{\phi}_{,\mu}
+\mathsfbi{\beta}^{{}_T}\mathsfbi{\alpha}_{,\mu}
-\mathsfbi{M}^{{}_T}\mathsfbi{\sigma} w_{\mu}=0,
\label{potential 4}
\end{align}
\end{subequations}
where $\mathsfbi{\sigma}^{{}_T}\equiv\partial\mathcal{L}_\mathcal{F}/\partial (\tilde{\mathfrak{D}}\mathsfbi{\phi})$, $\mathsfbi{\beta}^{{}_T}\equiv\mathsfbi{\sigma}^{{}_T}\partial \mathsfbi{C}/\partial (\mathfrak{D}\mathsfbi{\alpha})$\if0\cite{momentum}\fi. Taking the four-gradient of $\mathcal{L}_\mathcal{F}(\tilde{\mathfrak{D}}\mathsfbi{\phi})$ and combined with Eq. (\ref{potential 4}) yields
\begin{equation}
\mathsfbi{M}^{{}_T}\mathsfbi{\sigma}w_{\mu,\nu}w^{\nu}={\mathcal{L}_\mathcal{F}}_{,\mu},
\label{relativistic w eq.}
\end{equation}
which describes the dynamics of $\mathbf{w}$. Using a normalized vector $\mathbf{u}(\equiv \mathbf{w}/|\mathbf{w}|)$ ($u^{\alpha}u_{\alpha}=1$), the energy-momentum tensor is given by
\begin{equation}
\begin{split}
T^{\mu}{}_{\nu}
=\frac{\partial\mathcal{L}_\mathcal{F}}{\partial\mathsfbi{\phi}_{,\mu}}\mathsfbi{\phi}_{,\nu}
+\frac{\partial\mathcal{L}_\mathcal{F}}{\partial\mathsfbi{\alpha}_{,\mu}}\mathsfbi{\alpha}_{,\nu}
-\mathcal{L}_\mathcal{F}\delta^{\mu}_{\nu}
=\mathsfbi{M}^{{}_T}\mathsfbi{\sigma}
|\mathbf{w}|^2u^{\mu}u_{\nu}-\mathcal{L}_\mathcal{F}\delta^{\mu}_{\nu}.
\end{split}
\label{T}
\end{equation}
From the shift symmetry in space-time, we have $T^{\mu\nu}{}_{,\nu}=0$ which is equivalent to Eq. (\ref{relativistic w eq.}). Now $\mathsfbi{M}^{{}_T}\mathsfbi{\sigma}|\mathbf{w}|^2$ and $\mathcal{L}_\mathcal{F}$ act as the inertial mass density and the pressure respectively; namely the total mass density is to be defined by $\rho \equiv \mathsfbi{M}^{{}_T}\mathsfbi{\sigma}|\mathbf{w}|^2-\mathcal{L}_\mathcal{F}$. Multiplying Eq. (\ref{phi eq. 4}) by $\mathsfbi{M}$ yields 
\begin{equation}
(\mathsfbi{M}^{{}_T}\mathsfbi{\sigma}w^{\mu})_{,\mu}=(\rho_0 u^{\mu})_{,\mu}=0,
\label{rest mass}
\end{equation}
which may be interpreted as the conservation of the rest-mass density $\rho_0$. Eq. (\ref{relativistic w eq.}) is rewritten as the equation for $\mathbf{u}$:
\begin{equation}
\left\{(\rho+\mathcal{L}_\mathcal{F})u_{\mu}u^{\nu}\right\}_{,\nu}
=\mathcal{L}_\mathcal{F}{}_{,\mu}.
\label{U eq.}
\end{equation}

\subsection{Symmetry breaking and interactions}
Like in any well established classical field theory, interactions are incorporated, in principle, at the action level. Here we follow the same principle, and discuss some possible interactions which are allowed in our framework. As previously mentioned, the free fluid $\mathcal{L}_\mathcal{F}(\tilde{\mathfrak{D}}\boldsymbol{\phi})$ possesses shift symmetries of $\boldsymbol{\phi}$ and $\boldsymbol{\alpha}$ associated with its convective currents. Therefore, interactions may appear as the non-conservation of convective currents caused by the symmetry breaking.

A simple example of such interaction on $\boldsymbol{\phi}$ may be given by $\mathcal{L}=\mathcal{L}_\mathcal{F}(\tilde{\mathfrak{D}}\mathsfbi{\phi})+\mathcal{U}(\mathsfbi{\phi})$, which breaks the shift symmetry of $\boldsymbol{\phi}$ and breaks the rest-mass conservation as well. Yukawa-type interaction $\mathcal{U}\propto\bar{\psi}\psi \phi$ ($\phi$ is a single scalar for simplicity) is also possible that converts the rest-mass energy of the fluid to energy of the Dirac field $\psi$. Another interesting example may be the rest-mass exchange between fluids; in a 2-fluid system given by $\mathcal{L}=\mathcal{L}_a(\tilde{\mathfrak{D}}\phi_a)+\mathcal{L}_b(\tilde{\mathfrak{D}}\phi_b)+\mathcal{U}(\phi_a,\phi_b)$ (each fluid sector $\mathcal{L}_I$ ($I=a,b$) have a single-component kernel $\tilde{\mathfrak{D}}\phi_I$ containing $\mathbf{w}_I$, $\boldsymbol{\alpha}_I$, and $M_I$), one can observe the exchanging of their rest-mass density. Especially when $\mathcal{U}(\phi_a,\phi_b)=\mathcal{U}(\phi_a-\phi_b)$ with $M_a=M_b=M$, the total rest-mass current $\rho_{0a}\mathbf{u}_a+\rho_{0b}\mathbf{u}_b$ conserves.

Gauge interaction on $\boldsymbol{\alpha}$ is also naturally considered, where we see the exchange of non-Abelian charge via gauge interaction. Let $\mathsfbi{\alpha}$ be a $SU(n)$ multiplet, and impose invariance of the Lagrangian density under $SU(n)$ transformations of $\mathsfbi{\alpha}$, while $\mathsfbi{\phi}$ and $\mathsfbi{M}$ are just multiple-component real scalars. In order to incorporate gauge symmetry into the theory, the associated kernel (\ref{kernel 4}) itself must be invariant under local-gauge transformations. This implies on the one hand, quadratic dependance on $\mathfrak{D}\mathsfbi{\alpha}$, i.e. $\mathsfbi{C}(\Upsilon)$ with $\Upsilon \equiv (\mathfrak{D}\mathsfbi{\alpha})^{\dagger}\,\mathfrak{D}\mathsfbi{\alpha}$ (The conjugate $\boldsymbol{\beta}$ becomes also a multiplet; $\boldsymbol{\beta}=\partial\mathsfbi{C}^{{}_T}/\partial\Upsilon\ \mathsfbi{\sigma}\mathfrak{D}\boldsymbol{\alpha}$). And on the other, that derivatives at kernel are to be modified according to $\mathfrak{D}(=w^{\nu}\partial_{\nu} \rightarrow w^{\nu}\mathcal{D}_{\nu})$, where $\mathcal{D}_{\mu}\equiv\partial_{\mu}-ie\mathsfbi{t}_aA^a_{\mu}$ is the gauge-covariant derivative ($e$, $\mathsfbi{t}_a$, and $A^a_{\mu}$ are the coupling constant, generator, and gauge field, $a = 1,2,\cdots,n^2-1$) breaking the shift symmetry of $\boldsymbol{\alpha}$. The total fluid-gauge coupled system is given by
\begin{equation}
S=\int\mathcal{L}_\mathcal{F}(\tilde{\mathfrak{D}}\boldsymbol{\phi})\mathrm{d}^4x
+S_\mathrm{Y\!M}
\label{total action}
\end{equation}
with $\tilde{\mathfrak{D}}\mathsfbi{\phi}=\mathfrak{D}\mathsfbi{\phi}
-\frac{1}{2}\mathsfbi{M}w_{\alpha}w^{\alpha}+\mathsfbi{C}(\Upsilon)$. The gauge coupling in the Clebsch term does not affect the conservation law (\ref{phi eq. 4}), so that the rest-mass conservation (\ref{rest mass}) again holds. $SU(n)$-charge current arises from the gauge symmetry:
\begin{equation}
\begin{split}
J^{\mu}_a&=-ie\left(\mathsfbi{\beta}^{\dagger}\,\mathsfbi{t}_a\mathsfbi{\alpha}
-\mathsfbi{\alpha}^{\dagger}\,\mathsfbi{t}_a\mathsfbi{\beta}\right)w^{\mu}\\
&=-ie|\mathbf{w}|\frac{\partial \mathsfbi{C}^{{}_T}}{\partial \Upsilon}\boldsymbol{\sigma}\left((\mathfrak{D}\mathsfbi{\alpha})^\dagger\,\mathsfbi{t}_a\mathsfbi{\alpha}
-\mathsfbi{\alpha}^{\dagger}\,\mathsfbi{t}_a\mathfrak{D}\mathsfbi{\alpha}\right)u^{\mu}.
\end{split}
\label{current}
\end{equation}
Taking the four-gradient of $\mathcal{L}_\mathcal{F}(\tilde{\mathfrak{D}}\mathsfbi{\phi})$ yields
\begin{equation}
\left\{\left(\rho+\mathcal{L}_\mathcal{F}\right)u_{\mu}u^{\nu}\right\}{}_{,\nu}
=\mathcal{L}_\mathcal{F}{}_{,\mu}-F^a_{\mu\nu}J^{\nu}_a,
\label{CFD eq.}
\end{equation}
where $F^a_{\mu\nu}\equiv A^a_{\nu,\mu}-A^a_{\mu,\nu}+ef^a{}_{bc} A^b_{\mu}A^c_{\nu}$ is the field strength, $f^{abc}$ is the structure constants of the Lie algebra: $[\mathsfbi{t}_a,\mathsfbi{t}_b]=if_{abc}\mathsfbi{t}^c$.

To appreciate the reach and consequences of our formalism, it is instructive to contrast our picture with pioneering, and current, works \cite{Bis03, Jackiw04,Nair12,Nair14,Nair16}.

\noindent ($i$) Our $\boldsymbol{\phi}$ ($m$ real scalars) and $\boldsymbol{\alpha}$ ($SU(n)$ complex-scalar multiplet) amount to $m+2n$ real scalars, which reduces to $1+2n$ ($m=1$) degrees of freedom when the fluid carries only mass and charge. In contrast, the previous works include the group element $g \in SU(n)$ and mass which amount to $n^2$ real scalars \cite{SU2}. Let us also note that the present formalism enables us to consider the entropy current when $m=2$, while it is not obvious how such extension could be implemented from these works.

\noindent ($ii$) It is remarkable that (\ref{CFD eq.}) is obtained solely based on convection and gauge symmetry, whereas previous works on the other hand must rely on further assumptions. One being, the dynamics of $SU(n)$ charge on the basis of Wong's equation \cite{Wong70}, which is an estimated model of $SU(n)$-charged classical particle. And another, the continuum limit of Wong's particle by replacing particle label as the Lagrangian coordinate, guaranteeing the covariant conservation of charge current. The charge conservation relies on Wong's model and continuum limit; the framework is not a selfconsistent one, as it has roots in a particular model of classical-particle mechanics.



\noindent ($iii$) In the present formalism, the natural symplectic structure is derived via the co-tangent bundle of the configuration space $\{\boldsymbol{\phi},\boldsymbol{\alpha}\}$. This constitutes an issue for previous approaches, which starts from the particle-substratum Lagrangian density $\frac{1}{2}\rho v^2-U$ (or Hamiltonian density $\frac{1}{2}\rho v^2+U$) \emph{provided} Poisson brackets; the symplectic structure is \emph{artificially} introduced so that the Hamiltonian density $\frac{1}{2}\rho v^2+U$ yields the Euler equation and mass conservation \cite{Jackiw04,Nair12,Nair16}.

\noindent ($iv$) As a final remark, a similar treatment for the $U(1)$-charged fluid trivially follows by imposing $U(1)$ symmetry on a complex scalar $\alpha$. On the other hand, such a $U(1)$-gauge coupling model does not seem feasible in these formulations where $g=e^{i\theta}$ and $A_\mu$ give the only free variables which are not enough to describe electrically-charged fluid and the gauge field in its usual sense. Indeed the resultant Clebsch parametrization reads $u_\mu\propto\theta_\mu-eA_\mu$, where the vortical motion is completely constrained by $A_\mu$ due to the defect of the degrees of freedom \cite{irregular U1}. 

\section{Conclusion}
In this paper, we have reformulated the theory of the perfect fluid based on the symmetry behind the convective current. Peculiarly, its Lagrangian density $\mathcal{L}_\mathcal{F}(\tilde{\mathfrak{D}}\mathsfbi{\phi})$ is uniquely characterized by the quantity $\tilde{\mathfrak{D}}\mathsfbi{\phi}$, which we termed as kernel, composed by dynamical fields ($\mathsfbi{\phi}$, $\mathsfbi{\alpha}$), velocity $\mathbf{v}$ and constants $\mathsfbi{M}$. A minimal model for rotational flow follows once provided the kernel (\ref{kernel 3}) (or kernel (\ref{kernel 4}) in the relativistic case) with $m=n=1$, so the action contains only $(m+n=)2$ fields $\phi$ and $\alpha$. By discarding the particle-substratum approach built upon $\mathcal{L}_0=\frac{1}{2}\rho v^2-U$ and thereby Lin constraints of conventional formulations, we succeeded at obtaining an, impossible otherwise, Lagrangian density in configuration space, which is actually equivalent to abandon the conventional equations for non-observable potentials such as $\phi$, $\psi$, and $\alpha$ due to the presence of the Clebsch-potential energy $U_C$. Not only in such dimensional reduction, but the configuration-space formalism has also an advantage of much wider applicability. In case when the inverse functions $\mathsfbi{F};\boldsymbol{\sigma}\mapsto\tilde{\mathfrak{D}}\boldsymbol{\phi}$ and $\mathsfbi{H};\boldsymbol{\beta}\mapsto\mathfrak{D}\boldsymbol{\alpha}$ do not exist, the symplectic structure cannot be formulated, where the canonical phase-space formalism fails. The simplest example of this may be an incompressible fluid, where the Poisson brackets do not exist in the regular manner of \cite{Jackiw04,Nair12,Nair16}. Even in such case, the configuration-space formalism does hold, yielding the proper fluid equations consistently.



In our formulation, $\boldsymbol{\phi}$ and $\boldsymbol{\alpha}$ are not just secondary quantities parametrizing velocity $\mathbf{u}$, but dynamical fields capable of interacting with itself as well as other fundamental fields, playing a more active role in describing observable physics. As an example of the latter, we mentioned a possible scenario with a Yukawa coupling on $\boldsymbol{\phi}$, causing the exchange of the energy contained in the rest-mass density $\rho_0$ with the Dirac field. Regarding models with self-interactions, we discussed scenarios where $\phi$'s actively trigger the direct exchange of the rest mass. The non-Abelian (Abelian) gauge symmetry on $\boldsymbol{\alpha}$ is also possible naturally resulting in a non-Abelian (Abelian) Lorentz force on fluid, which provides a \emph{pure gauge theory of the perfect fluid} for the first time. Discarding the particle-substratum paradigm our construction does not require any further assumptions other than gauge symmetry as in the usual field theories. In an analogous way, the diffeomorphism covariance imposed on $\boldsymbol{\alpha}$ leads to a pure diffeomorphism-gauge theory of the spin fluid, again, free from extra assumptions. Also $\boldsymbol{\alpha}$ of the diffeomorphism covariance gives minimum degrees of freedom to describe the spin-gravity coupling in completely novel way. This will be reported in a follow-up publication \cite{MorAri}.

The field theoretic picture enables us to reconsider an underlying relation between rotational flow and the Clebsch potentials. According to Helmholtz's vortex theorem \cite{Lamb}, the vortex can be neither created nor destroyed without an external force; namely the creation and annihilation of vortex is finally attributed to a force imposed on the fluid. As long as we take the action principle the starting point, the external force is introduced as an interaction term in the Lagrangian; among various interactions of this field theory, the electromagnetic field and its extension (such as Yang-Mills field) may be naturally considered as an external force on the fluid. These gauge fields have spin, so their emission and absorption is accompanied by the exchange of angular momentum, causing the vortex as the localized orbital angular momentum of fluid. Thus, the presence of the Clebsch potentials allows to couple with these gauge forces that produce vorticity. The Clebsch potentials serve as the \emph{mediator} to the external field causing vorticity.

Finally let us mention about field quantization. Some pioneers discussed about the quantum field theory of fluid, where excitation of Lagrangian position of fluid element is the target of quantization \cite{Endlich11,GS15}. Note that the Lagrangian picture of the fluid implies the rest-mass conservation; each world line of a fluid element running from the past to future does not terminate at some points. The present formalism, on the contrary, is formulated without such picture, and is even able to describe the creation and annihilation of the rest mass. This suggests the possibility of another quantum field theory of very different type, where the fluid itself may be obtained as an excitation from the vacuum due to interactions with matter and gauge particles.


\subsection*{Acknowledgement}
T. A. currently belongs to Institute of Material and Systems for Sustainability, Nagoya University, Japan. The present work is supported by the Japan Society for the Promotion of Science for Young Scientists.\\

\appendix
\section{Derivation of Eq. (\ref{H of convection})}\label{derivation of Hc}
For the first step, we write the time derivative of the first term $\rho_\mathrm{c}\partial_t\phi$ in terms of convective derivative:
\begin{equation*}
\mathcal{H}_\mathrm{c}
=\rho_\mathrm{c}\mathfrak{D}\phi_\mathrm{c}-\rho_\mathrm{c}v^j\phi_{,j}-\mathcal{L}_\mathrm{c}\circ F_\mathrm{c}(\rho_\mathrm{c})
\end{equation*}
where the first term may be rewritten as $\rho_\mathrm{c}F_\mathrm{c}(\rho_\mathrm{c})$. Differentiation of $\rho_\mathrm{c}F_\mathrm{c}(\rho_\mathrm{c})-\mathcal{L}_\mathrm{c}\circ F_\mathrm{c}(\rho_\mathrm{c})$ by $\rho_\mathrm{c}$ yields
\begin{equation*}
1\times F_\mathrm{c}(\rho_\mathrm{c})+\rho_\mathrm{c} F'_\mathrm{c}(\rho_\mathrm{c})-\rho_\mathrm{c}F'_\mathrm{c}(\rho_\mathrm{c})=F_\mathrm{c}(\rho_\mathrm{c})
\end{equation*}
Thus $\rho_\mathrm{c}F_\mathrm{c}(\rho_\mathrm{c})-\mathcal{L}_\mathrm{c}\circ F_\mathrm{c}(\rho_\mathrm{c})$ can be expressed as $\int^{\rho_\mathrm{c}}F_\mathrm{c}(\xi)\mathrm{d}\xi$, where the constant is not written explicitly.


\begin{thebibliography}{99}
\bibitem{alphabeta}
Different notations are employed for the Clebsch potentials so that $(\alpha,\beta)$ form canonical-conjugate pairs.
\bibitem{UC}
Since $U_C$ contains only $\rho$, $s$, and $\beta$, it only changes the canonical equations of $\phi$, $\psi$ and $\alpha$ neither of which is observed.
\bibitem{SU2}
In case of $SU(2)$ coupling, requires $n^2=4$ which is less than our $2n+1=5$. However, $n=2$ without gauge coupling requires 4 scalars, while our formalism needs only 2. Apart from the $SU(2)$-gauge coupling case, the present formalism requires less degrees of freedom.
\bibitem{irregular U1}
In recent \cite{Nair14,Nair16} direct application is avoided, adding rest-mass current ($\boldsymbol{J}_m$) and charge current ($\boldsymbol{J}_e$) with their constitutive model $\boldsymbol{J}_m=\boldsymbol{J}_e$, containing rest-mass density, $\theta$, $\alpha$, $\beta$, and $g\in U(1)$, equal to 4 real scalars if the gauge freedom is unified into $\theta$, while ours contain only a real $\phi$ and a complex $\alpha$, which amounts to 3 real scalars.
\end{thebibliography}
\end{document}